\documentclass[conference]{IEEEtran}
\IEEEoverridecommandlockouts
\usepackage{cite}
\usepackage{tabularx,booktabs}
\usepackage{amsmath,amssymb,amsfonts}
\usepackage{algorithmic}
\usepackage{graphicx}
\usepackage{textcomp}
\usepackage{float} 
\newcolumntype{C}{>{\centering\arraybackslash}X}
\usepackage{lipsum}
\usepackage{hyperref}
\usepackage{booktabs}
\usepackage{multirow}
\def\BibTeX{{\rm B\kern-.05em{\sc i\kern-.025em b}\kern-.08em
    T\kern-.1667em\lower.7ex\hbox{E}\kern-.125emX}}
\begin{document}

\title{Towards Asynchronous Motor Imagery-Based Brain-Computer Interfaces: a joint training scheme using deep learning}

\author{\IEEEauthorblockN{Patcharin Cheng\IEEEauthorrefmark{1}, Phairot Autthasan\IEEEauthorrefmark{1}, Boriwat Pijarana\IEEEauthorrefmark{1}, Ekapol Chuangsuwanich\IEEEauthorrefmark{2} \\
and Theerawit Wilaiprasitporn\IEEEauthorrefmark{1}}
\IEEEauthorblockA{
\IEEEauthorrefmark{1}Bio-inspired Robotics and Neural Engineering Lab, \\School of Information Science and Technology, Vidyasirimedhi Institute of Science \& Technology, Thailand
\\Email: theerawit.w@vistec.ac.th
}
\IEEEauthorblockA{
\IEEEauthorrefmark{2}Computer Engineering Department, Chulalongkorn
University, Bangkok, Thailand.}
}

\maketitle

\begin{abstract}
In this paper, the deep learning (DL) approach is applied to a joint training scheme for asynchronous motor imagery-based Brain-Computer Interface (BCI). The proposed DL approach is a cascade of one-dimensional convolutional neural networks and fully-connected neural networks (CNN-FC). The focus is mainly on three types of brain responses: non-imagery EEG (\textit{background EEG}), (\textit{pure imagery}) EEG, and EEG during the transitional period between background EEG and pure imagery (\textit{transitional imagery}). The study of transitional imagery signals should provide greater insight into real-world scenarios. It may be inferred that pure imagery and transitional EEG are high and low power EEG imagery, respectively. Moreover, the results from the CNN-FC are compared to the conventional approach for motor imagery-BCI, namely the common spatial pattern (CSP) for feature extraction and support vector machine (SVM) for classification (CSP-SVM). Under a joint training scheme, pure and transitional imagery are treated as the same class, while background EEG is another class. Ten-fold cross-validation is used to evaluate whether the joint training scheme significantly improves the performance task of classifying pure and transitional imagery signals from background EEG. Using sparse of just a few electrode channels ($C_{z}$, $C_{3}$ and $C_{4}$), mean accuracy reaches 71.52\% and 70.27\% for CNN-FC and CSP-SVM, respectively. On the other hand, mean accuracy without the joint training scheme achieve only 62.68\% and 52.41\% for CNN-FC and CSP-SVM, respectively.


\end{abstract}

\begin{IEEEkeywords}
deep learning-based BCI, motor-imagery BCI, joint training, asynchronous BCI, EEG-based BCI 
\end{IEEEkeywords}

\section{Introduction}
The brain-computer interface (BCI) is a neural-inspired technology with the potential to become an alternative communication channel between man and machine. Bypassing peripheral nerves and muscles, users directly send commands to the machine through information presented in the brain signals. Electroencephalography (EEG) is a common tool for measuring brain activity in BCI applications \cite{1}. One type of brain signal is called Motor imagery (MI). To generate MI related-brain signals, a user is asked to imagine motor function execution such as hands or feet movements. Event-related desynchronization/synchronization (ERD/ERS) is a variation in brain signals during MI. ERD/ERS decoding has led to the development of an MI-based BCI \cite{2}. Recent studies on brain-controlled wheelchairs \cite{3} and quadcopters \cite{4} demonstrates that the possibilities for application of MI-based BCI. There are two major research approaches for improving the MI-EEG-based BCI. The first of which draws upon knowledge from behavioural neuroscience to enhance the MI-EEG signal-to-noise ratio (SNR). Instead of a simple hand movement imagery task, switching to a more complex one, such as writing Chinese characters, has been shown to enhance MI-EEG significantly \cite{7827070}. It may be concluded that high complex imagery tasks improve the SNR of MI-EEG. The second approach is via the development of more sophisticated classification algorithms using modern neural network techniques, namely deep learning (DL). Recently, DL approaches based on convolutional neural networks (CNN) and stacked auto-encoders (SAE) have shown promising results on MI-EEG classification \cite{1741-2552-14-1-016003}. In MI-EEG, the DL approach still has much room for improvement. Even though there are various studies on MI-EEG, the experimental protocols for collecting EEG signals leave a lot to be desired in terms of practicality. Almost all datasets have been gathered from synchronised tasks, allowing researchers to know the exact onset and offset time of imagery events \cite{4359220}. This bypasses many of the difficulties in signal cleaning and normalization. However, in real applications, the MI-based BCI needs to be asynchronous. This motivates the researchers to explore MI-EEG classification tasks in more practical scenarios.

This study mainly focuses on three types of brain response, namely non-imagery EEG (\textit{background EEG}), \textit{pure imagery} EEG, and EEG during the transitional period between pure imagery and background EEG (\textit{transitional imagery}). The study of transitional imagery signals should provide greater insight into real-world scenarios. It may be inferred that pure imagery and transitional EEG are high and low power imagery EEG signals, respectively. DL models employ a cascade of one-dimensional convolutional neural networks (CNNs) \cite{doi:10.1162/neco.1997.9.8.1735} and a fully-connected network (FC) or CNN-FC is used in this study. The experiments compare three classification tasks: Firstly, to classify pure imagery from background EEG, secondly, transitional imagery from background EEG, and lastly, to separate both pure and transitional imagery from background EEG using a joint training scheme, considering pure and transitional imagery signals as the same class. Experimental results demonstrate that a joint training scheme using pure and transitional imagery (high and low power imagery) can improve the performance of the model in separating both pure and transitional imagery from background EEG. The researchers believe that this exploratory work is the first step towards practical MI-EEG-based BCI applications dealing with transitional periods in the signal.

\section{METHODOLOGY}
This section firstly introduces a publicly available motor imagery EEG (MI-EEG) dataset, and then explains the segmentation and pre-processing of data. The proposed DL approach is a cascade of one-dimensional CNN-FC. Then we compare outcomes from CNN-FC to the conventional approach, namely common spatial pattern and support vector machine (CSP-SVM) \cite{10.3389/fnins.2013.00267}.

\begin{figure}[th]
\centering
\includegraphics[width=1.0\linewidth]{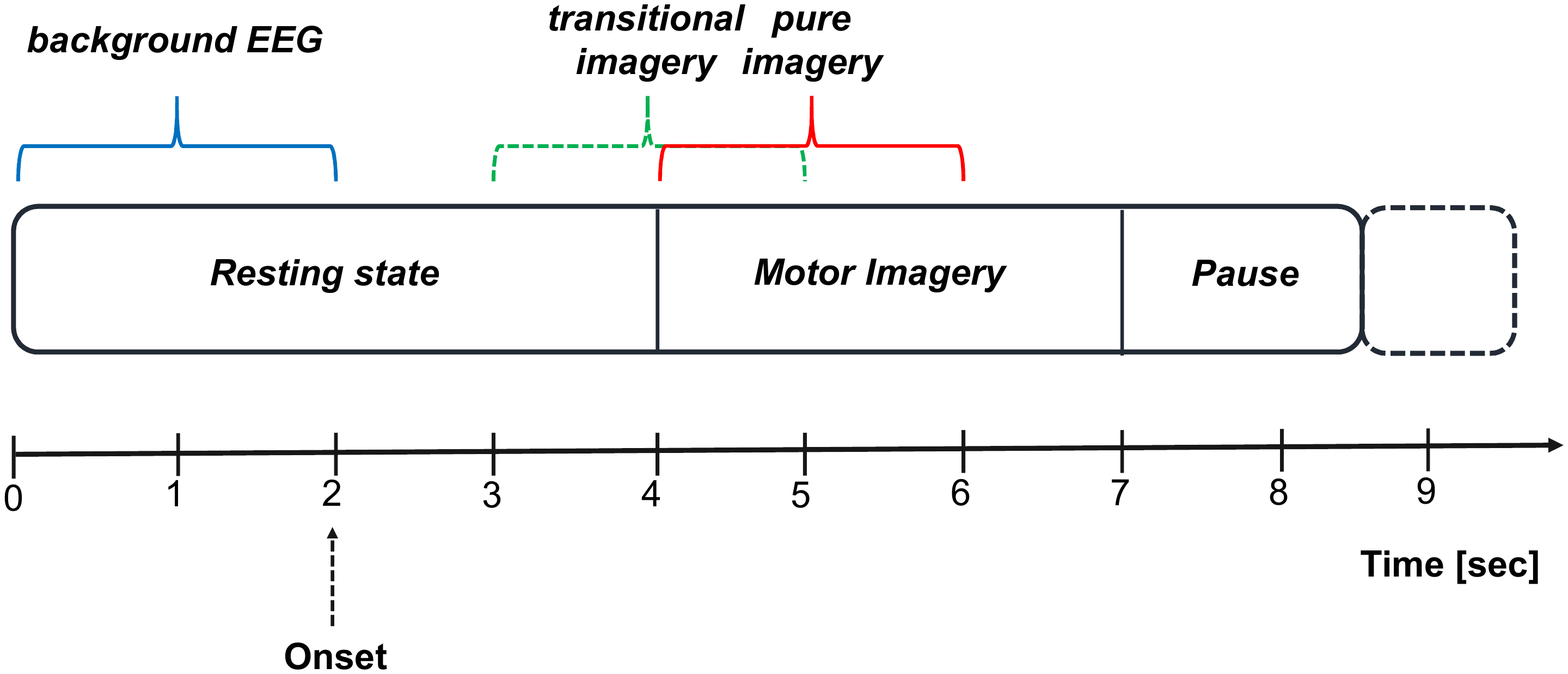} 
\caption{The collection protocol from MI-EEG. The first period (from 0 to 2 seconds) is denoted as background EEG. The second period is the transitional imagery interval (from 3 to 5 seconds). Finally, the interval of pure imagery is from 4 to 6 seconds.} 
\label{period of data}
\end{figure}

\begin{figure*}
\centering
\includegraphics[width=1\linewidth]{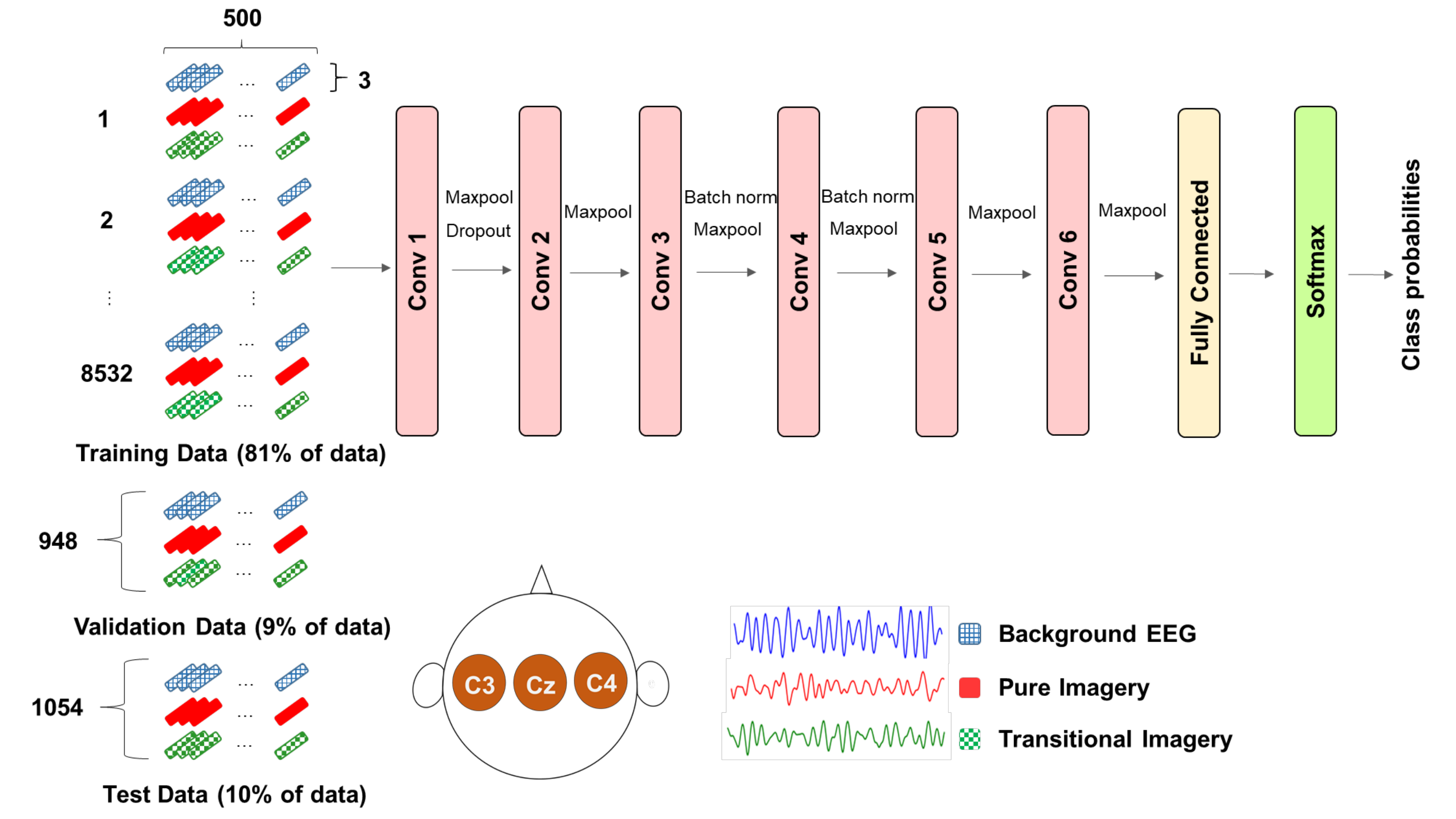}
\caption{Implementation of the cascade CNN-FC model on pre-processed data for training: talidation: testing samples are 81:9:10}
\label{structure}
\end{figure*}

\subsection{Dataset}
In this study, the MI-EEG dataset was used \cite{4359220}. Nine healthy subjects participated in five sessions of motor imagery experiments. The first two sessions were recorded without feedback and the other three sessions had combined online feedback. Participants were instructed to imagine movement on either their left or right hand. Each of the first two sessions consisted of 120 trials and the remaining three sessions 160 trials. As stated in the experimental protocol (\autoref{period of data}), one trial consisted of a time interval of between 8.5 to 9.0 seconds. The first interval is a resting state. During the following three-second period the participants were requested to carry out motor imagery tasks. Finally, another time interval of between 1.5 and 2 seconds consisted of a resting state lasting until the end of the trial. In summary, every subject had to take part in 280 trials, making this a substantial dataset in terms of both number of trials and participants. The data was selected and pre-processed using the following steps:
\begin{itemize}
\item EEG data with 250 Hz sampling frequency were extracted from three electrodes around the motor function area of a human brain ($C_{z}$, $C_{3}$ and $C_{4}$).
\item A Bandpass was filtered at 7.0--30.0 Hz.
\item As shown in \autoref{period of data}, the EEG signal was divided from a single trial into three classes, namely background EEG, transitional imagery, and pure imagery.
\end{itemize}


Moreover, those trials containing artifacts which were marked in the original dataset were removed. Hence, in the experiments, only clean or accepted trials were picked for further study. It is assumed that MI-EEG from clean trials could provide higher SNR and would be better in CNN-FC training. Finally, the MI-EEG data was arranged into the 0--1 range by standardization.

\subsection{Proposed Deep Learning Approach}
The architecture of the DL model is illustrated in \autoref{structure}. Three classes of pre-processed data (background EEG, pure imagery, and transitional imagery) from nine participants were prepared. As described in the data section, 12,600 samples $\times$ 500 points (two seconds-length) $\times$ 3 electrodes were obtained from each EEG class. According to Section II \textit{A}, only clean trials were selected from the total data, eventually resulting in 10,534 samples $\times$ 500 points $\times$ 3 electrodes for each class. 

The DL model began with a six layers of CNNs. The CNNs were used to extract spatial information and for learning the activation patterns of different MI-EEG signals. Thus, the shape of activation patterns (i.e. voltage values from different times) and their location (i.e. MI-EEG channel) were learnt from the stack of CNN layers. Outputs from the CNNs were then flattened into a single one-dimensional vector and subsequently fed into a stack of FCs. Finally, a softmax function was incorporated for binary classification.



\subsection{Experiments and Evaluation}
The purpose of this study is to answer the research question `` \textit{Can a joint training scheme help to improve the classification of both pure and transitional imagery EEG from background EEG ?} " The researchers considered that the pure and transitional imagery segments were from the same class, and background EEG was from another class. To answer the question, five experimental tasks were conducted as shown below:
\begin{itemize}
  \item Task 1: to distinguish pure imagery from background EEG 
  \item Task 2: to distinguish transitional imagery from background EEG
  \item Task 3: to distinguish both pure and transitional imagery (treated as the same class) from background EEG using a joint training scheme with transitional imagery: pure imagery (1:1)
  \item Task 4: to distinguish both pure and transitional imagery (treated as the same class) from background EEG using a joint training scheme with transitional imagery: pure imagery (1:4)
  \item Task 5: to distinguish both pure and transitional imagery (treated as the same class) from background EEG using a joint training scheme with transitional imagery: pure imagery (4:1)
\end{itemize}

Among the five tasks, the same training, validation, and testing procedure was performed (as shown in \autoref{structure}). Ten-fold cross-validation was used to train and evaluate each model. 81\% of the samples were used for the training, 9\% were used for the validation and the remaining 10\% were used for the testing.


\subsubsection{Deep Learning (DL) Approach}
As depicted in \autoref{structure}, the architecture of the model contains eight learnt layers--six convolutional and two fully-connected (CNN-FC), implemented using Keras \cite{chollet2015keras} with parameter configurations as follows:

\begin{itemize}
\item Six convolutional layers (CNNs) with 32, 64, 64, 128, 256, and 512 filters, respectively.
\item Kernel size of 3 for all layers.
\item The ReLU non-linearity was applied to the output of every convolutional and fully-connected layer.
\item Max-pooling with a pooling size of 2 was applied after each convolutional layer.
\item Dropout with a probability of 0.2 was used in the first convolutional layer.
\item Batch normalization was applied after the second and third convolutional layers.  
\item Output from the last convolutional layer was flattened before being taken as fully-connected input.
\item 256 neurons were in the first fully-connected layer (FC).
\item Softmax was applied for classification (FC).  
\item The optimizer was Adadelta with a learning rate of 1.0.
\end{itemize}

\subsubsection{Machine Learning (ML) Baseline}
To demonstrate the superiority in the proposed DL approach, namely the CNN-FC, a conventional ML for MI-BCI, namely common spatial pattern with support vector machine (CSP-SVM), was used for a baseline. Here, the CSP was implemented using the MNE-python package \cite{10.3389/fnins.2013.00267}, and configured using default parameters to obtain three spatial components as outputs. The outputs from CSP were then fed into the SVM. The SVM was implemented using Scikit-Learn \cite{scikit-learn} and the linear kernel function set with a \textit{C} value of 1. The CSP-SVM evaluated the performance using the same ten-fold cross-validation as in the CNN-FC, except for the ignorance of validation. Training and testing sets are generally sufficient for evaluating the ML approach.

Testing accuracy, sensitivity, and specificity from CNN-FC and CSP-SVM were used to compare the results of both approaches among the five experimental tasks. For statistical analysis of the experimental results, one-way repeated measures analysis of variance (ANOVA) was performed, based on the assumption of sphericity. Greenhouse-Geisser correction was used when the data violates the sphericity assumption. In post hoc analysis, Bonferroni correction and pairwise comparison was performed.


\section{Results}
\begin{figure}
\centering
\includegraphics[width=0.9\linewidth]{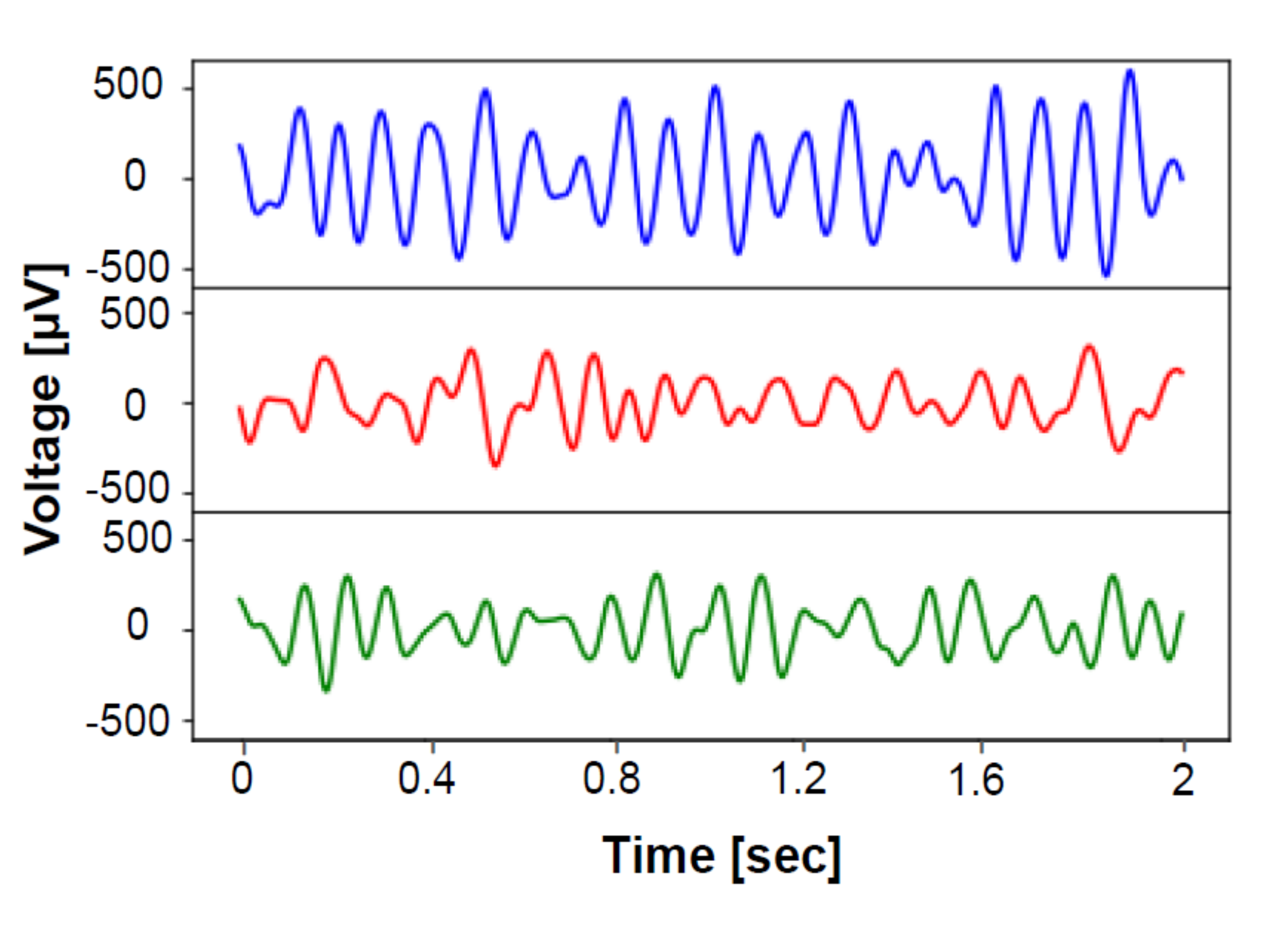} 
\caption{An example of qualitative comparison among single trial EEG from EEG background (top), pure imagery (middle), and transitional imagery (bottom). The presented EEGs (from $C_{3}$ electrode) were pre-processed using a bandpass filter at 7.0-30.0 Hz.}
\label{g222}
\end{figure}

\autoref{g222} shows an example of a single trial EEG background (top), pure imagery (middle), and transitional imagery (bottom) signals. Presented EEGs (from $C_{3}$ electrode) were pre-processed by a bandpass filter at 7.0-30.0 Hz. These signals can not be classified by human inspection, but its can be classified by the proposed DL approach and the ML baseline.

\begin{figure}
\centering
\includegraphics[width=1\linewidth]{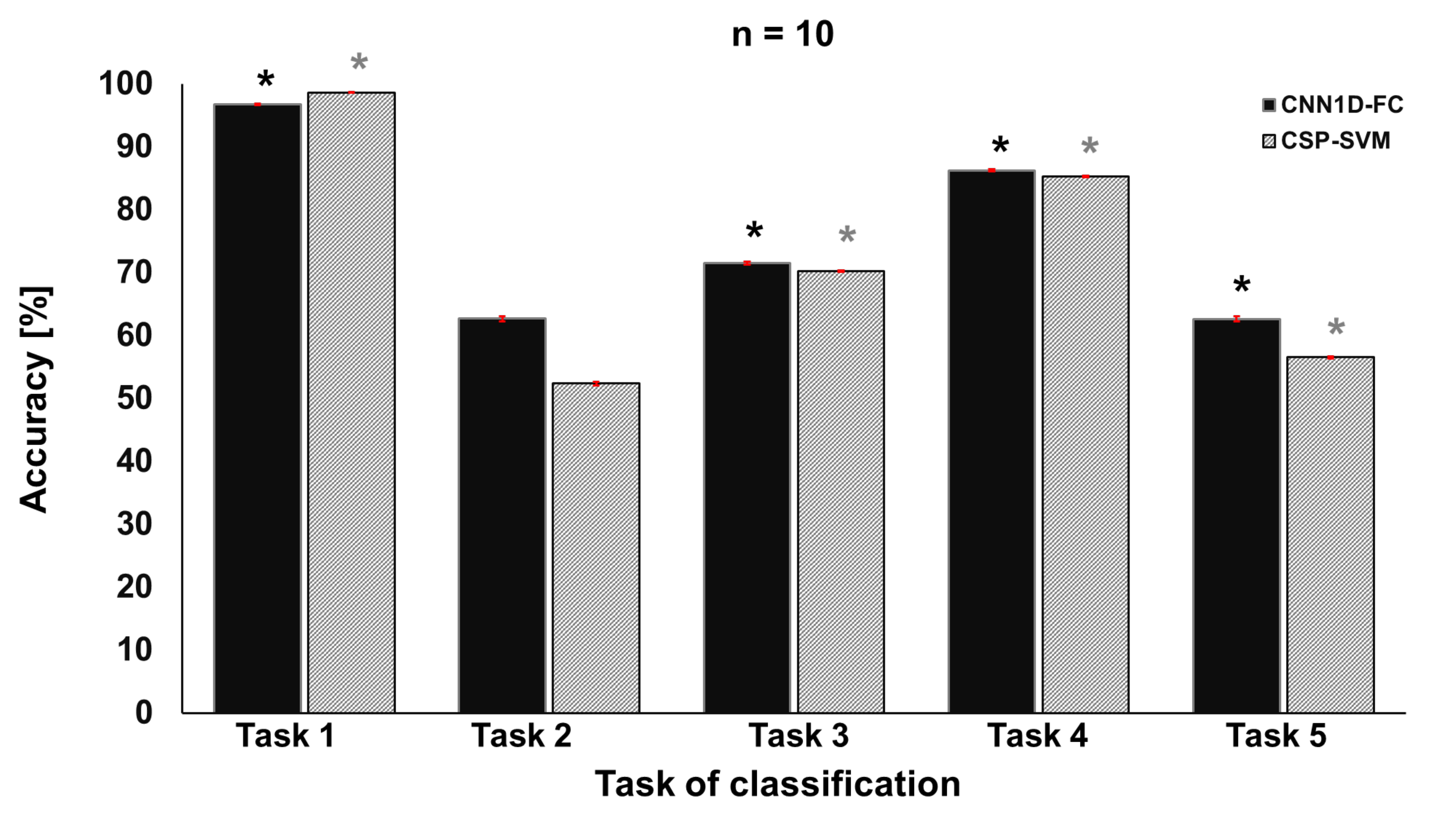} 
\caption{Comparison of mean accuracies (n = 10) between different tasks. The error bar is standard. 
* denotes that Task 1 is significantly higher than the others, p$<$0.01.}
\label{g_all}
\end{figure}

The results from the five experimental tasks are shown in \autoref{g_all}. Task 1 was designed to distinguish pure imagery from background EEG. Task 2 was designed to distinguish transitional imagery from background EEG. Task 3 was designed to distinguish balanced number of training samples between both pure and transitional imagery (treated as the same class) and background EEG using a joint training scheme with a ratio of 1:1. Task 4 was designed to distinguish imbalanced number of training samples between both pure and transitional imagery (treated as the same class) and background EEG using a joint training scheme with a ratio of 1:4. Finally, Task 5 was designed to distinguish imbalanced number of training samples between both pure and transitional imagery (treated as the same class) and background EEG using a joint training scheme with a ratio of 4:1. One-way repeated measures ANOVA with Greenhouse-Geisser correction reported significant differences among the experimental tasks, classified by CNN-FC and CSP-SVM approaches, \emph{F(1.899,17.089) = 2384.927, p$<$0.01}, \emph{F(1.628,14.652) = 11994.267, p$<$0.01}, respectively. In pairwise comparison of CNN-FC, the mean accuracy from Task 1 was significantly higher than the other tasks because only one pure imagery and EEG background were contained in the classification. In the same way, Task 4 was the second highest of the five tasks. This result was caused by the ratio of pure imagery:transitional imagery being 4:1. However, for Tasks 3 and Task 5 (with a joint training scheme and pure imagery:transitional imagery the sampling ratios were 1:1 and 1:4, respectively), giving significantly higher mean accuracies than Task 2 (without a joint training scheme). This answers the research question as to whether a joint training scheme can improve classification tasks between low power imagery EEG (transitional imagery) and background EEG. In the CSP-SVM approach, pairwise comparison results on all experimental tasks were consistent with those using the CNN-FC approach.  
\autoref{my-label} presents a comparison between mean accuracies, sensitivities, and specificities among five experimental tasks using CNN-FC and CSP-SVM approaches. The standard t-test, assuming unequal variances, was performed to compare the results between the CNN-FC and CSP-SVM. The CNN-FC gave significantly higher accuracy in Tasks 2 to Task 5 (\emph{t(17)=2.11, p$<$0.05}, \emph{t(11)=2.2.20, p$<$0.05}, \emph{t(16)=2.12, p$<$0.05} and \emph{t(13)=2.16, p$<$0.05}, respectively). The results in bold shown in Table 1 are significantly higher than the others. In summary, the mean accuracy results from the CNN-FC are promising and significantly higher than those from the CSP-SVM in Tasks 2 to Task 5.

\begin{table*}[ht]
\centering
\caption{Comparison of mean accuracy, sensitivity, and specificity in binary classification (mean±standard error) among five experimental tasks using two implemented approaches for classification. The numbers in bold are statistically significantly higher than the others, p$<$0.01} 
\label{my-label}
\resizebox{1.5\columnwidth}{!}{
  \begin{tabular}{@{}ccccccc@{}}
  \toprule
  \multirow{2}{*}{\textbf{Task}} & \multicolumn{2}{c}{\textbf{Accuracy {[}\%{]}}} & \multicolumn{2}{c}{\textbf{Sensitivity {[}\%{]}}} & \multicolumn{2}{c}{\textbf{Specificity {[}\%{]}}} \\ \cmidrule(lr){2-3}\cmidrule(l){4-5} \cmidrule(l){6-7}  
                                 & \textbf{CNN-FC}      & \textbf{CSP-SVM}      & \textbf{CNN-FC}        & \textbf{CSP-SVM}       & \textbf{CNN-FC}        & \textbf{CSP-SVM}       \\ \midrule
  \textbf{Task 1}                & 96.72 $\pm$ 0.12               & \textbf{98.65 $\pm$ 0.03}      & 96.39 $\pm$ 0.38                  &\textbf{98.44 $\pm$ 0.06}       & 97.04 $\pm$ 0.35                 &\textbf{98.86 $\pm$ 0.00}       \\ \midrule
\textbf{Task 2}                & \textbf{62.68 $\pm$ 0.38}       & 52.41 $\pm$ 0.30              &\textbf{68.03 $\pm$ 2.02}         & 52.01 $\pm$ 0.34                & 57.32 $\pm$ 2.45         & 52.81 $\pm$ 0.37               \\ \midrule
\textbf{Task 3}                & \textbf{71.52 $\pm$ 0.29}       & 70.27 $\pm$ 0.09              &\textbf{59.17 $\pm$ 2.83}         & 52.50 $\pm$ 0.15                & 83.87 $\pm$ 3.04                  & 88.03 $\pm$ 0.15       \\ \midrule
\textbf{Task 4}                & \textbf{86.27 $\pm$ 0.19}      & 85.32 $\pm$ 0.13               & 76.26 $\pm$ 0.61                 &\textbf{78.56 $\pm$ 0.16}       &\textbf{96.28 $\pm$ 0.44}         & 92.09 $\pm$ 0.21               \\ \midrule
\textbf{Task 5}                & \textbf{62.63 $\pm$ 0.41}       & 56.56 $\pm$ 0.20               & \textbf{64.12 $\pm$ 2.19}         & 45.92 $\pm$ 0.43                & 61.14 $\pm$ 2.54                  & \textbf{67.19 $\pm$ 0.37}       \\ \bottomrule
  \end{tabular}
  }
\end{table*}

\section{Discussion}
The experimental results in \autoref{g_all} demonstrate that a joint training scheme can improve the performance of the model in the separation of pure and transitional imagery (high and low power imagery) from background EEG. The study of transitional imagery signals offers greater insight into real-world scenarios. Moreover, the mean accuracy of CNN-FC reaches 86.27\% in the best joint training scheme Task 4, and this could be a good starting point for the proposed approach.

In a practical scenario, an online MI-based BCI does not have onset and offset trigger events. Thus, the computer or machine will never know the start and end point of motor imagery signals from EEG \textbf{(asynchronously)}. Moreover, the incoming EEG will also be fed to the system continuously (and randomly). This is the reason why the study of transitional imagery recognition is very important to online system development. So far, none of the research groups have investigated this issue. It is hoped that the findings in this study will inspire further BCI research.

The DL approach (CNN-FC in this study) uses a joint training scheme and not only shows MI-based BCI improvement but how this fundamental concept can also lead to further applications (for example, by adopting the DL approach to medical images). Furthermore, the DL approach with a combination of joint training and transfer learning is currently active in various applied data-driven research. However, it is still a challenge to incorporate these schemes and approaches in MI-based BCI.




\section{Conclusions and Future Works}
In summary, the proposed DL approach (a cascade of one-dimensional CNNs and FCs) was used with a joint training scheme for classification of background EEG (resting/normal state) and motor imagery (MI) EEG. The joint training scheme would benefit the development of an asynchronous MI-based brain-computer interface (BCI) in real-world scenarios. An asynchronous MI-based BCI should have the capability to recognize both pure imagery and non-pure imagery EEG (transitional imagery in this study). According to the experiments in this study, the joint training scheme outperformed typical training. In the future, the researchers propose to focus on adjusting the DL model to improve the performance even further and incorporate the approach into online MI-based BCI applications.

\bibliographystyle{unsrt}
\bibliography{Reference.bib}
\end{document}